\documentstyle[12pt]{article}

\newcommand{\al}{\alpha} 
\newcommand{\G}{\Gamma}

\begin{document}
\thispagestyle{empty}
\begin{flushright}
MZ-TH/98-59\\
December 1998\\
\end{flushright}
\vspace{0.5cm}
\begin{center}
{\Large\bf On the running electromagnetic coupling constant
at $M_Z$}\\[1truecm]
{\large J.G.~K\"orner$^1$, A.A.~Pivovarov$^{1,2}$ and 
K.Schilcher$^1$}\\[.1cm]
$^1$ Institut f\"ur Physik, Johannes-Gutenberg-Universit\"at,\\[-.1truecm]
  Staudinger Weg 7, D-55099 Mainz, Germany\\[.2truecm]
$^2$ Institute for Nuclear Research of the\\[-.1truecm]
  Russian Academy of Sciences, Moscow 117312
\vspace{0.6truecm}
\end{center}

\begin{abstract}
We present a discussion on how to 
define the running electromagnetic coupling constant
at $M_Z$ or some other intermediate scale as e.g.
$m_\Upsilon$. We argue that a natural definition consistent
with general requirements of the
renormalization group should be based on  
Euclidean values of the momentum of the photon propagator
as the appropriate scale.
We demonstrate in an explicit 
example of evaluating the running coupling constant 
at the scale of the $\Upsilon$ resonance
mass 
that the usual definition of the 
hadronic contribution with a principal value
prescription
is inconsistent. 
In the determination of the value of 
$\al$ at $M_Z$ the 
numerical difference due to using a Euclidean definition 
rather than the principal value one is 
comparable in size to the errors
caused by existing experimental
and QCD inputs to the evaluation of 
$\al(M_Z)$.
\end{abstract}
\newpage

In applications to high precision tests of the standard model \cite{SMrev}
with observables near the $Z$ boson peak 
the electromagnetic coupling constant should be used at a scale 
of the order of the $Z$ boson mass
$M_Z$, (see e.g. \cite{CERNwork,Hollik}).
The running  
electromagnetic coupling constant at $M_Z$
has even been chosen
as a standard reference parameter \cite{PDG}. 
It differs  numerically from the value of the fine structure 
constant $\alpha^{-1}=137.036\ldots$ defined at 
zero momentum or from the Coulomb law for heavy nonrelativistic
particles. The change is usually 
accounted for through the renormalization group
equation \cite{RG,GellMannLow}.
Because the fine structure constant is defined
at vanishing momentum and is taken as initial value in the solution of
the renormalization group equation, the 
running  
electromagnetic coupling constant at $M_Z$
is an infrared sensitive quantity in as much as 
the contribution of strong interaction is not easy to compute
due to the 
nonperturbative region at small energies. Therefore this contribution
is usually taken into account in the leading order of
electromagnetic interaction within a semi-phenomenological
approximation 
through a dispersion relation.
There has been a renewal of interest in 
a precise determination of the
hadronic
contribution during the last years in particular in connection with
the constraints on the Higgs boson mass \cite{Higgs}. 
Some recent references giving
a state-of-the-art analysis of this contribution
are
\cite{Davier1,Schilcher,Kuhn,Davier}. 
A quasi-analytical approach was used in 
\cite{Krasnikov} where some references to earlier paper can be found
(see also \cite{ITEP,burkhardt}).
An extremely thorough 
data-based analysis is given in \cite{Jegerlehner}.

In the present note we critically discuss the
definition of the running electromagnetic coupling constant
at $M_Z$ as it is used in the literature.
The standard approach consists in using the principal value
prescription at the appropriate scale in the physical domain on the positive
energy semiaxis.
We argue that a natural definition consistent
with general requirements and the standard notion of running
used 
in renormalization group applications should be based on the 
Euclidean momentum of the photon propagator
as the appropriate scale.

The running coupling $\al(q^2)$
is defined through the (one-photon irreducible) photon vacuum 
polarization function
$\Pi_{\gamma}(q^2)$ as 
\begin{equation}
  \label{running}
\al(q^2)={\al\over 1-\Pi_{\gamma}(q^2)}  .
\end{equation}
$\Pi_{\gamma}(q^2)$
contains
both leptonic and hadronic contributions. The hadronic part
of the polarization function $\Pi_{\gamma}(q^2)$
(with one subtraction at zero 
momentum) reads
\begin{equation}
  \label{disp}
\Pi_{\gamma}^{had}(q^2)=-\frac{\al}{3\pi}\,q^2\!\!
\int\limits_{4 m_\pi^2}^\infty{R_h(s)ds\over s(s-q^2-i0)}
\end{equation}
where $R_h(s)$ is the normalized cross section of $e^+e^-$ annihilation
into hadrons. 
Let us introduce, for convenience, 
the polarization function  $\Pi(q^2)$
\begin{equation}
  \label{dispBare}
\Pi(q^2)
= -q^2 \int_0^\infty{R(s)ds\over s(s-q^2-i0)}
\end{equation}
such that
\begin{equation}
  \label{dispRen}
\Pi_{\gamma}(q^2)=\frac{\al}{3\pi}\Pi(q^2)
\end{equation}
where $R(s)$ is the corresponding spectral density.
Note that $\al(q^2)$ is defined for every complex value of $q^2$ by
eq.~(\ref{disp}). For real negative $q^2$ the polarization function 
$\Pi_{\gamma}(q^2)$ (and $\Pi(q^2)$ as well)
is a positive real number 
because the spectral density $R(s)$
is positive.

The definition (\ref{running}) is used in renormalization 
group applications and the scale $q^2$
is taken to be a real negative number which corresponds to 
a propagator in the Euclidean domain.
The Euclidean definition is mainly 
used in applications of grand unified theories \cite{GUT},
supersymmetry at large energy \cite{SUSY}, physics at the Planck's scale
etc. 

For the precise study of
the physics at the $Z$ boson pole
the effective electromagnetic interaction coupling constant
$\bar \al$
is represented in the form
\begin{equation}
  \label{fitform}
\bar\al={\al\over 1-\Delta \al}\, .
\end{equation}
Numerically one obtains a real positive number 
for $\Delta \al$.

A theoretical expression for $\Delta \al$
is defined in the present literature
directly on the positive semiaxis by 
making use of the principal value
prescription
for the singularity of the integrand in eq.~(\ref{disp})
\begin{equation}
  \label{standdef}
\Delta\al= {\rm Re}\,\Pi_{\gamma}(M_Z^2),\qquad  
{\rm Re}\,\Pi_{\gamma}(M_Z^2)=-\frac{\al}{3\pi}M_Z^2 \,
{\rm P}\!\!\!\int\limits_{4 m_\pi^2}^\infty{R(s)ds\over s(s-M_Z^2)}.
\end{equation}
Here ${\rm P}\!\int$ denotes the principal value of the integral. 
This makes $\Delta\al$ real (the initial $i0$ prescription
for the integral gives it an imaginary part)
which is appropriate for a coupling constant.
We argue that this prescription is not adequate for the physical
situation at hand
and does not
correspond to the notion of a running coupling used in the   
standard
renormalization group applications. The latter corresponds to scales taken 
in the Euclidean domain
\begin{equation}
  \label{eucldef}
\al_E(\mu^2)={\al\over 1-\Pi_{\gamma}(-\mu^2)},\qquad  
\Pi_{\gamma}(-\mu^2)=\frac{\al}{3\pi}\,\mu^2 \!\!\!
\int\limits_{4 m_\pi^2}^\infty{R(s)ds\over s(s+\mu^2)}.
\end{equation}
The
running  
electromagnetic coupling constant at the scale $M_Z$ 
is defined then as $\al_E(M_Z^2)$.
Therefore, for the phenomenological parameter 
``running  
electromagnetic coupling constant at the scale $M_Z$''
denoted by $\bar \al$ we have two representations:

\noindent i) the standard one with a principal value prescription
\begin{equation}
  \label{apv}
\bar \al=\al_{PV}(M_Z^2), \qquad  \Delta\al=  {\rm Re}\,\Pi_{\gamma}(M_Z^2)
\end{equation}
ii) the alternative one in the Euclidean domain
\begin{equation}
  \label{aeucl}
\bar \al=\al_E(M_Z^2), \qquad  \Delta\al= \Pi_{\gamma}(-M_Z^2)
\end{equation}
which is defined through
\begin{equation}
  \label{euclres}
\al_E(M_Z^2)={\al\over 1-\Pi_{\gamma}(-M_Z^2)},\qquad  
\Pi_{\gamma}(-M_Z^2)=\frac{\al}{3\pi}\,M_Z^2\!\!\!
\int\limits_{4 m_\pi^2}^\infty{R(s)ds\over s(s+M_Z^2)}.
\end{equation}

We suggest that the Euclidean version is used.
The idea of changing scales is embodied
in the renormalization group equation which allows one
to control large logarithms. Therefore theoretically
one is dealing with a logarithm of the ratio of two scales.  
Note that the notion of scale becomes rather imprecise
as soon as complex numbers are involved.
For example, numerically $M_Z^2$ has the same 'scale' as 
$e^{i\pi}M_Z^2=-M_Z^2$.
The choice of an appropriate scale is determined by the particular 
kinematics and the higher order corrections of each
particular process in question.
In the leading logarithmic approximation, however,
keeping the finite corrections to large logarithms
is beyond the accuracy of the 
approximation that renders all 
scales with the same absolute value equivalent.
As a reference value for the coupling constant the usual choice
is to take a Euclidean point.
This problem was discussed for strong interactions in \cite{early}
where different versions of a real part or absolute value 
definition of the coupling
constant
at complex points have been also studied.
In the general case it is difficult to decide 
on how to deal with observables including complex numbers
within the renormalization group resummation of logarithms.
For two-point functions, however, there is a natural solution to this problem
based on their analytic properties given by 
the dispersion representation
\cite{Tau,b4}.  

Below we discuss these two possibilities
of defining the ``running  
electromagnetic coupling constant at the scale $M_Z$''.

First we show that the two definitions (Euclidean
and principal value)
are close numerically for
applications in the vicinity of the $Z$ boson peak 
discussed in the literature.
Let us take eq.~(\ref{disp}) and split the whole 
region of integration into two parts separated by $s_0$
\begin{equation}
  \label{split}
\Pi(q^2)=-q^2\int_0^{s_0}{R(s)ds\over s(s-q^2-i0)}
-q^2\int_{s_0}^\infty {R(s)ds\over s(s-q^2-i0)}.
\end{equation}
If $|q^2|$ is chosen such that 
$|q^2|\gg s_0$ one can expand the
denominator in the first integral. Then, if $s_0$ is large enough one
can use perturbation theory for the spectral density in the second integral.
For illustrative purposes we choose a very simplified
approximation for $R(s)$, namely 
$R(s)={\rm const}=1$ for  $s>s_0$.
Then one obtains
\begin{equation}
  \label{result}
\Pi(q^2)=\int_0^{s_0}{R(s)\over s}ds
+\ln{|s_0-q^2|\over s_0}      
\end{equation}
where the principal value
prescription has been used.
Expanding eq.~(\ref{result})
in the limit $|q^2|\gg s_0$ 
one finally obtains
\begin{equation}
  \label{resultfinal}
\Pi(q^2)=\int_0^{s_0}{R(s)\over s}ds
+\ln{|q^2|\over s_0}
\end{equation}
which is independent of the phase of the complex number $q^2$.
The same result can be obtained also directly from eq.~(\ref{dispBare})
in this limit.
Therefore in the above approximation with the suggested regime of
variables
both the Euclidean
and principal value definitions are equivalent
numerically.
Later on we discuss corrections to this leading order approximation
which depend on whether the Euclidean
and principal value definition is used.
  
This is a qualitative picture. Because the above simplifying
assumptions
can be expected to correctly embody 
the main features of a more sophisticated numerical analysis 
it is clear that the numerical change stemming from the choice of
$q^2$ in the Euclidean domain 
rather than using the conventional definition 
is under control at the scale of $M_Z$
and does not jeopardize current phenomenology.
However,
the definition in the Euclidean domain given in eq.~(\ref{euclres})
is preferable from a theoretical point of view.
It is natural. It gives a real number. It is smooth. It is consistent
with the renormalization group.  

And the principal value
prescription has equally obvious deficiencies.
It is ad hoc. It gives a real part of a propagator which 
is not directly 
related to a coupling constant in a renormalization group sense. 
It is not smooth.

The last deficiency is, in fact, the most crucial one.
Let us present more details.
We take the principal value
definition at $q^2=M^2_Z$
and compute the polarization function for a model spectral density 
$R(s)=\theta(s-s_0)$ obtaining
\begin{equation}
  \label{nonsmo}
\Pi(M^2_Z)=\ln{|s_0-M^2_Z|\over s_0}\, .
\end{equation}
Note first that 
the polarization function (\ref{nonsmo}) 
gives a rather curious result
\begin{equation}
  \label{curiuos}
P\!\!\!\!\int\limits_{M_Z^2/2}^\infty{ds\over s(s-M_Z^2)}=0
\end{equation}
which means that the contribution of all states 
with masses larger than $M_Z/\sqrt{2}\sim 60~{\rm GeV}$
is exactly equal to zero assuming the asymptotic spectral density in this
region to be a constant. Also there is  a sign change 
in the vicinity of $M_Z/\sqrt{2}$.
This feature persists for any realistic $R(s)$ in the vicinity of some
point $s^*\simeq M_Z^2/2$ because in this region QCD perturbation theory
works well and the spectral density is rather smooth and close to its
asymptotic value which is almost a constant (up to a slow logarithmic
decrease).
Therefore one gets the exact equality
\begin{equation}
  \label{morecuriuos}
P\int\limits_{s^*}^\infty{R(s)ds\over s(s-M_Z^2)}=0
\end{equation}
for some $s^*\sim M_Z^2/2$.

Furthermore, if one takes $s_0=M^2_Z$ in eq.~(\ref{nonsmo})
then the logarithm is ill-defined. This can be seen to be 
a consequence of the principal value definition 
eq.~(\ref{standdef}).
Even if this is a rather academic example we nevertheless
take it as a warning
(because there is 
no sharp increase of the spectral density
or changes in general in the vicinity of the 
$Z$ boson mass).
More realistic situations are considered below. 
In contrast to the principal value definition 
the Euclidean definition is fine also in this case
\begin{equation}
  \label{smo}
\Pi(-M^2_Z)=\ln{s_0+M^2_Z \over s_0}\, .
\end{equation}
The reason for the ill-defined behavior
of eqs.~(\ref{nonsmo}) and 
(\ref{standdef})
is clear. The principal value prescription leads to 
a distribution
$P\frac{1}{x}$ which is defined only on smooth functions.
A product of two distributions
\begin{equation}
  \label{nonintmod}
\left( P{1\over s- M^2_Z}\right)\theta(s- M^2_Z)
\end{equation} 
is not an integrable function.
An ad hoc definition with a principal value
prescription fails to define a value
for the running coupling at some particular points 
and one has to introduce further 
rules for such cases.

Also in a more realistic situation one needs 
the running electromagnetic coupling constant at the scales around 
the masses of resonances of the $J/\psi$ or $\Upsilon$ families
to account for their leptonic widths. With the principal value
definition 
it is impossible to compute 
the running electromagnetic coupling constant at the scales around 
the resonance  masses. Indeed, the correction $\Delta \al$
to the running electromagnetic coupling constant at the scale of the 
$\Upsilon$ meson mass $m_\Upsilon$ is given by an ill-defined integral of
the product of two distributions
\begin{equation}
  \label{nonintres}
\left( P{1\over s - m^2_\Upsilon}\right)\delta(s - m^2_\Upsilon).
\end{equation} 
This quantity is not defined as a distribution
for the same reason as a product of two distributions
is not defined
when their singular points coincide.

Leaving the mathematical statement about the 
ill-defined behavior of a product of two 
distributions aside, in practice 
(for finite widths of the resonances or for some sharp but still smooth
increase of the spectral function) the results following from the
principal 
value
definition will be unstable.
As an explicit example we take the spectral density corresponding
a single
Breit-Wigner resonance and calculate its contribution
to $Re \Pi(M_Z^2)$.
The Breit-Wigner spectral function 
is given by 
\begin{equation}
  \label{BW}
R_{BW}(s)=\frac{1}{\pi}{\G M\over (s-M^2)^2+\G^2 M^2}
\end{equation}
where $M$ and $\G$ are the mass and the width of the
resonance. As $\G\rightarrow 0$ one obtains 
$R_{BW}(s)\rightarrow \delta(s- M^2)$.
After integration (with proper care for the point $s=0$)
one finds
\begin{equation}
  \label{BWint}
M_Z^2 
{\rm P}\!\!\!\int^\infty{R_{BW}(s)ds\over s(s-M_Z^2)}
= {M^2-M_Z^2\over (M^2-M_Z^2)^2+\G^2 M^2}+\ldots=
{\Delta\over \Delta^2+\G^2 M^2}+\ldots
\end{equation}
with $\Delta=M^2-M_Z^2$. Only potentially 
singular terms for the limit $\G\rightarrow 0$
in the vicinity 
$M^2\sim M_Z^2$ have been kept.
The last function in eq. (\ref{BWint})
has its extremal points at  $\Delta=\pm \G M$  
with the values
\begin{equation}
  \label{extval}
\left.{\Delta\over \Delta^2+\G^2 M^2}\right|_{\Delta=\pm \G M} 
=\pm \frac{1}{2\G M}.   
\end{equation}
There is no regular limit $\G\rightarrow 0$ 
and further rules would be
required
to deal with this limit.
Note that the result (\ref{BWint}) can be obtained without 
explicit
integration of the Breit-Wigner spectrum.
Since the Breit-Wigner function
can be regarded as a regularization of a $\delta$-distribution 
one can introduce
a regularization of $P\frac{1}{x}$ in the class of infinitely
smooth function instead. An example is 
\[
\lim_{\epsilon\to 0}{x\over x^2+\epsilon^2}=P\frac{1}{x}\qquad
{\rm and } \qquad
\lim_{\epsilon\to 0}{(s-M_Z^2)\over (s-M_Z^2)^2+\epsilon^2}
=P{1\over s-M_Z^2}\,
.
\]
Then after the integration with an infinitely narrow resonance one gets
\begin{equation}
  \label{regpv}
M_Z^2 
\int{\delta(s-M^2)(s-M_Z^2)ds\over s[(s-M_Z^2)^2+\epsilon^2]}
= {M_Z^2\over M^2}
{\Delta\over \Delta^2+\epsilon^2}={\Delta\over \Delta^2+\epsilon^2}+\ldots
\end{equation}
for $M^2 \sim M_Z^2$.
The regularization can not be unambiguously removed,
i.e. there is no unique limit at $\epsilon=0$ in the vicinity 
of $\Delta=0$.
Of course, this is a reflection of the fact that the product of two 
distributions is ill-defined. 

We will not dwell on the ill-defined behavior when a $\theta$-function type 
spectral density is used. This is a
realistic situation when one computes the light
quark contributions to the running electromagnetic coupling constant
normalized in the vicinity of a sharp raise of the spectral
density around $1.5~{\rm GeV}^2$.

With a Euclidean definition none of the above difficulties
appear.
Also because in this case the polarization function is defined in
the Euclidean domain one need not integrate all data 
(only a small region near the origin requires explicit 
integration) and can use all 
the power
of perturbation theory (e.g. \cite{ChetKuehn}).
Although 
special care has to be taken about the subtraction at zero momentum
that enters the definition of the coupling constant
and makes it an infrared sensitive 
quantity. 
For this purpose, however, more sophisticated 
means can be used that will increase the accuracy \cite{svz,Nasrallah,p0}.

We add two further remarks.
The first concerns the leptonic contribution.
For a lepton with the mass $m_l$ the asymptotic form of this
contribution reads
 \begin{equation}
  \label{lep}
\Pi_{lept}=\ln\left(\frac{M_Z^2 }{m_l^2}\right)- \frac{5}{3} 
\end{equation}
and has the same real part for any real phase $\varphi$ of $e^{i\varphi}
M_Z^2 $.
When the asymptotic form is used both prescriptions are equivalent numerically.

The second remark is related to the higher order contributions of 
the electroweak
interactions. In the next order of the electroweak interaction there is a
contribution of the $Z$ boson peak to 
the polarization function $\Pi_{\gamma}(q^2)$
due to $\gamma Z$ transitions (e.g. \cite{gZ}).
Therefore in that order one has 
to interpret the product of the principal value distribution with the
sharp Breit-Wigner spectrum of the
$Z$ boson pole itself.
Again this problem is not present within the Euclidean definition.

Now we discuss very briefly the numerical difference that can result 
from the change of the definition
of the running coupling constant at $M_Z$.
Our model for the hadronic spectral density
$R(s)$ is simple and is 
mainly designed for illustrative purposes such that one can easily 
trace the difference 
between the principal value and Euclidean definitions of the running 
coupling constant.
The spectral density is chosen such 
that all calculations can be done analytically which is 
convenient for the purpose of estimating the order of
magnitude of 
the difference in the two definitions.
For the light quarks $u$, $d$, $s$ we assume the existence of a low lying
resonance (like $\rho$, $\omega$, $\varphi$) and a continuum.
In general, we take the following form
of the spectrum
for every light quark flavor 
\[
R_{light}(s)=3Q_q^2[ 2m^2\delta(s-m^2)+\theta(s-2m^2)]
\]
according to the model of ref.~\cite{FESR1}
with $Q_q$ being a light quark fractional charge. 
The coupling of the low lying resonances have been replaced by the
duality interval $2m^2$.
For heavy quarks we take the simplest model of the form
\[
R_{heavy}(s)=3Q_Q^2 \theta(s-4m_Q^2)
\]
which represents the partonic asymptotic value with a naive
step function
two-quark-threshold with $Q_Q$ being a heavy quark charge.
Collecting everything together we find the following results.
The three light quarks give the result for $\Pi(q^2)$
in the general form
\[
\Pi_{light}(q^2)=2\left(
{-2 q^2\over m^2-q^2-i0}+\ln
{|2m^2 -q^2|\over 2m^2}\right).
\]
Expanding at $|q^2|\sim M_Z^2$ one gets
\begin{equation}
  \label{lightcont}
\Pi_{light}=2\left[2+\ln{|M_Z^2|\over 2m^2}+O(m^6/M_Z^6)\right]  
\end{equation}
which gives the same answer for both definition 
with a high precision. The difference starts at the order $O(m^6/M_Z^6)$
and is completely negligible for light resonance masses
of order $1~{\rm GeV}$ ($\rho$-, $\omega$-, $\varphi$-resonances, for instance).
For the charm contribution we find
\[
\Pi_{charm}^{PV}=\frac{4}{3}\ln{M_Z^2-4m_c^2\over 4m_c^2}
\]
in the case of the principal value prescription.
In the case of the Euclidean prescription one has
\[
\Pi_{charm}^{E}=\frac{4}{3}\ln{M_Z^2+4m_c^2\over 4m_c^2} \, .
\]
Expanding these formulae in the small ratio $4m_c^2/M_Z^2$ one finds
in the leading order 
\[
\Pi_{charm}^{PV,E}=
\frac{4}{3}\ln{M_Z^2\over 4m_c^2}\left(1\mp \frac{4m_c^2}{M_Z^2}
\frac{1}{\ln{M_Z^2\over 4m_c^2}}\right)\, .
\]
We keep this form for further numerical comparison
in the case of $b$ and $t$ quarks.
For the $b$ quark contribution one gets
\[
\Pi_{bottom}^{PV,E}
=\frac{1}{3}\ln{M_Z^2\over 4m_b^2}\left(1\mp \frac{4m_b^2}{M_Z^2}
\frac{1}{\ln{M_Z^2\over 4m_b^2}}\right),
\]
and for the $t$ quark
\[
\Pi_{top}^{PV,E}=\frac{4}{3}\ln{\left(1\mp {M_Z^2\over 4m_t^2}\right)}.
\]
For numerical estimates we take
$\sqrt{2} m= 1~{\rm GeV}$ for the light quark resonances,
$m_c= 1.4~{\rm GeV}$, $m_b= 4.8~{\rm GeV}$,
$m_t= 175~{\rm GeV}$, $M_Z= 91~{\rm GeV}$.
Note that the exact definition of the quark mass parameters 
is not required here because it is 
far beyond the accuracy of our simple model. Even more, these
parameters can be
considered as effective parameters serving to describe 
integrals over the threshold regions of quark production.
Nevertheless we stick to almost canonical values for the pole quark masses. 
While the absolute value of the contribution
will be obtained rather approximately,
the model however is fairly sufficient 
for our main purpose to estimate the difference
between the two definitions.
Numerically, eq.~(\ref{lightcont}) leads to the light quark 
contribution 
\[
{\rm light~quarks~(u~,d~,s) }=2\times(2+2\times 4.5)=22.0
\]
with both prescriptions.
For the contributions of the heavy quarks we find
\begin{eqnarray}
  \label{numHeavy}
{\rm c-quark}&=&\frac{4}{3}\times 7.0(1.\mp 0.14\times 10^{-3})\nonumber
\\
{\rm b-quark} &=&\frac{1}{3}\times 4.5(1.\mp 2.5\times 10^{-3})\nonumber
\\
{\rm t-quark}&=& \frac{4}{3}\times (\mp 68\times 10^{-3}).
\end{eqnarray}
Summing everything together one obtains
\[
22.0+\frac{4}{3}\times 7.0(1.\mp 0.14\times 10^{-3})_c
+\frac{1}{3}\times 4.5(1.\mp 2.5\times 10^{-3})_b
+\frac{4}{3}\times (\mp 68\times 10^{-3})_t
\]
\begin{equation}
  \label{hadronic}
=32.8 + 0.1\, \delta  \approx 33 + 0.1\, \delta
\end{equation}
where $\delta=- 1$ for the principal value definition
and $\delta= 1$ for  
the Euclidean definition.
One sees that the difference is saturated by the top quark
contribution.
This is natural because it has a mass closest to $M_Z$. 
Its contribution is small in absolute value but is completely different for
the two definitions. 

The lepton contribution
is taken into account according to the asymptotic 
formula eq.~(\ref{lep}) 
with $m_e=0.5~{\rm MeV}$, $m_\mu=0.1~{\rm GeV}$,
$m_\tau=1.8~{\rm GeV}$ which gives
\[
(24.2)_e+(13.6)_\mu+(7.8)_\tau-5=40.6 \approx 41 \, .
\]
The final result for the total contribution
of the charged fermions to the polarization function reads
\[
33 + 0.1\, \delta+41=74+0.1\, \delta \, .
\]
Dividing by $3\pi$ one obtains the total contribution to the inverse coupling
constant 
\[
(74+0.1\,\delta)/3\pi=7.9 + 0.01\, \delta \, .
\]
For the inverse running 
electromagnetic coupling constant we obtain 
\[
137.0-(7.9 + 0.01\, \delta)
=129.1-0.01\, \delta \, .
\]
Even though the central value of our approximate evaluation
is rather close to the results of more precise evaluations 
\cite{Davier1,Schilcher,Kuhn,Davier} this agreement 
should not be taken 
too seriously.
Our estimate is rather rough and serves the
purpose well to obtain the numerical change between 
the two definitions
of 
the running electromagnetic coupling constant
at $M_Z$.
At present the change 
due to the different definitions has no influence on current
phenomenology. 
This change is within 
the error bars for the uncertainty of the more precise values
$128.93\pm 0.06$ \cite{Schilcher}
or even with smaller errors
$128.93\pm 0.015_{\rm exp}\pm 0.015_{\rm th}$ 
\cite{Davier1,Kuhn,Davier}.
Therefore the Euclidean definition, 
which we consider more consistent 
theoretically, does not violate current phenomenology.
However, the change is comparable in size with the 
present uncertainties and when 
the experimental data used in the determination
of the running electromagnetic coupling constant
at $M_Z$ improves in the future
the difference between the two definitions
will become significant.

Our last numerical example concerns the order of magnitude
of the singular term in the coupling normalized
at the scale $m_\Upsilon$ within the principal value prescription.
With the same normalization as in our 
simple model 
the contribution of the $\Upsilon$-resonance to the spectral density 
in the Breit-Wigner approximation
reads
\begin{equation}
  \label{Upsilon}
R_{\Upsilon}(s)=\frac{2}{3} m_\Upsilon \Delta_\Upsilon 
R_{BW}(s, m_\Upsilon,\Gamma)
\end{equation}
where $\Delta_\Upsilon\approx 1~{\rm GeV}$ is its duality interval in energy
units \cite{FESR0}
related to its leptonic decay width, 
$\Delta_\Upsilon=27\pi \Gamma(\Upsilon\rightarrow e^+e^-)/2\alpha^2$ 
with $\Gamma(\Upsilon\rightarrow e^+e^-)=1.32~{\rm keV}$
while
$\Gamma = 52.5~{\rm keV}$ is its
full width.
The singular contribution
of the $\Upsilon$-resonance to the polarization function
abruptly changes between the two extremes
taken from eq.~(\ref{extval}) 
\[
\pm \frac{2}{3} m_\Upsilon \Delta_\Upsilon 
\left(\frac{1}{2 \Gamma m_\Upsilon}\right)
\]
as the normalization point passes the position of the resonance.
Numerically one gets
\[
\frac{\Delta_\Upsilon}{3\Gamma} = 6.7\times 10^3 
\]
which is far too big from the point of view of phenomenology.

The examples presented in this paper
clearly demonstrate the inconsistency of the present 
definition of the running electromagnetic coupling constant
within the principal value prescription. 
These problems are not noticeable, however, when 
one discusses a normalization point around $M_Z$
because the hadronic spectral density is smooth in this region.

We are going to present the results of an 
accurate numerical analysis within 
the Euclidean definition elsewhere.

To conclude, 
we suggest that the Euclidean domain 
definition of the running electromagnetic coupling constant
at $M_Z$ is used as a reference parameter 
for high precision tests of the standard model
at the $Z$ boson peak. It is free of the shortcomings
of the present definition 
which is based on the propagator at the physical value of 
the $Z$ boson mass within
the principal value prescription.

\noindent{\large \bf Acknowledgements}\\[2mm]
The work is supported in part
by Volkswagen Foundation under contract
No.~I/73611. 
A.A.~Pivovarov is  
supported in part by
the Russian Fund for Basic Research under contracts Nos.~96-01-01860
and 97-02-17065.
The present stay of A.A.~Pivovarov in Mainz was made possible 
due to Alexander von Humboldt fellowship.

\end{document}